\documentclass[a4paper,11pt]{article}
\usepackage{pos}
\usepackage{siunitx}
\usepackage{subcaption}
\usepackage{hyperref}
\usepackage{lineno}

\DeclareSIUnit \h {\ensuremath{\mathit{h}}}
\DeclareSIUnit \pc {pc}
\DeclareSIUnit \Mpc {Mpc}
\DeclareSIUnit \kpc {kpc}
\DeclareSIUnit \erg {erg}
\DeclareSIUnit \deg {deg}
\DeclareSIUnit \year {yr}
\DeclareSIUnit \Gyear {Gyr}
\DeclareSIUnit \gauss {G}
\DeclareSIUnit \EeV {EeV}
\DeclareSIUnit \PeV {PeV}
\DeclareSIUnit \tons {t}
\DeclareSIUnit \sr {sr}
\DeclareSIUnit\GB{GB}
\DeclareSIUnit\arcmin{'}
\DeclareSIUnit\degsymbol{\degree}

\title{Event reconstruction using pattern spectra and convolutional neural networks for the Cherenkov Telescope Array}
\ShortTitle{Event reconstruction using pattern spectra and CNNs for CTA}

\author*[a,b]{J. Aschersleben}
\author[a]{M. Vecchi}
\author[b]{M. H. F. Wilkinson}
\author[a]{R. F. Peletier}

\affiliation[a]{Kapteyn Astronomical Institute\\
University of Groningen, PO Box 800, 9700 AV Groningen, The Netherlands}

\affiliation[b]{Bernoulli Institute for Mathematics, Computer Science and Artificial Intelligence\\
University of Groningen, PO Box 407, NL-9700 AK Groningen, The Netherlands}

\forColl{CTA} 

\emailAdd{j.j.m.aschersleben@rug.nl}
\emailAdd{r.f.peletier@rug.nl}
\emailAdd{m.vecchi@rug.nl}
\emailAdd{m.h.f.wilkinson@rug.nl}

\abstract{The Cherenkov Telescope Array (CTA) is the future observatory for ground-based imaging atmospheric Cherenkov telescopes. Each telescope will provide a snapshot of gamma-ray induced particle showers by capturing the induced Cherenkov emission at ground level. The simulation of such events provides camera images that can be used as training data for convolutional neural networks (CNNs) to differentiate signals from background events and to determine the energy of the initial gamma-ray events. Pattern spectra are commonly used tools for image classification and provide the distributions of the sizes and shapes of features comprising an image. The application of pattern spectra on a CNN allows the selection of relevant combinations of features within an image. \\
In this work, we generate pattern spectra from simulated gamma-ray images to train a CNN for signal-background separation and energy reconstruction for CTA. We compare our results to a CNN trained with CTA images and find that the pattern spectra-based analysis is computationally less expensive but not competitive with the purely CTA images-based analysis. Thus, we conclude that the CNN must rely on additional features in the CTA images not captured by the pattern spectra.}

\FullConference{%
  7th Heidelberg International Symposium on High-Energy Gamma-Ray Astronomy (Gamma2022)\\
  4-8 July 2022\\
  Barcelona, Spain\\}

\begin{document}
\maketitle

\section{Introduction}
\noindent
Gamma rays induce air showers when interacting with the Earth's atmosphere, which results in the creation of secondary particles. These secondary particles can travel with a velocity larger than the speed of light in air, which induces a flash of Cherenkov light~\cite{Cherenkov1934}. The Cherenkov light contains information about the initial particle, such as its type, energy and direction, and can be captured by imaging atmospheric Cherenkov telescopes (IACTs) from the ground~\cite{de_Naurois_2015}. The Cherenkov Telescope Array (CTA)\footnote{www.cta-observatory.org}~\cite{CTA2019} is the future observatory for ground-based imaging atmospheric Cherenkov telescopes and will be located in the northern hemisphere at the Roque de los Muchachos Observatory in La Palma (CTA North) and in the southern hemisphere in the Atacama Desert in Chile (CTA South). It will consist of three different telescope types: Small-Sized Telescopes (SSTs), Medium-Sized Telescopes (MSTs) and Large-Sized Telescope (LSTs). The different telescope types will allow CTA to cover a wide energy range between $\SI{20}{\GeV}$ and $\SI{300}{\TeV}$ with an energy resolution of $\SI{5}{\percent}$ around $\SI{1}{\TeV}$ and an angular resolution of $\sim \SI{1}{\arcmin}$ on individual photons for the upper end of the CTA energy range~\cite{Gueta_2021}. \\
Gamma-ray events can be reconstructed by applying Hillas parameters~\cite{Hillas1985} on machine learning algorithms, such as Random Forest~\cite{albert2008} or Boosted Decision Trees~\cite{Ohm2009, Becherini2011}. Other reconstruction methods are based on semi-analytical or Gaussian photosphere shower models, such as model analysis~\cite{de_Naurois_2003} and 3D model analysis~\cite{LEMOINEGOUMARD2006195}. Lately, convolutional neural networks (CNNs) have been applied to IACT data~\cite{Jacquemont2021, Aschersleben2021, Nieto2019, Miener2021b, spencer2021deep} motivated by their success in other image classification and regression tasks~\cite{alzubaidi2021review}. However, one of the main drawbacks of this method is that the training of CNNs is computationally very expensive~\cite{strubell2019energy}. \\
In this work, we extract pattern spectra from CTA data aiming to reduce the computational resources needed to train a CNN for signal-background separation and gamma-ray energy reconstruction. Pattern spectra are commonly used tools for image classification~\cite{Urbach2007, batman1997size} and provide a 2-dimensional histogram of the sizes and shapes of features within an image. Our pattern spectra algorithm is based on the work presented in Urbach et al. (2007)~\cite{Urbach2007} and makes use of connected operators, which merge regions within an image with the same grey scale value. Compared to other pattern spectra algorithms, our method provides the following main advantages: (i) the computing time for creating the pattern spectra is independent of its dimensions and (ii) it is significantly less sensitive to noise~\cite{Urbach2007}.

\section{Dataset}
\noindent
Two different datasets have been used for the signal-background separation and the energy reconstruction, respectively. The signal-background dataset consists of $\sim 6 \cdot 10^5$ gamma-rays and $\sim 6 \cdot 10^5$ protons generated in a view cone of $\SI{10}{\degsymbol}$. The energy reconstruction dataset consists of $\sim 2.5 \cdot 10^6$ gamma-ray events generated with a $\SI{0.4}{\degsymbol}$ offset from the telescope pointing position. All events have been simulated for the southern CTA array with a telescope zenith angle of $\SI{20}{\degsymbol}$ (North pointing) and an energy range between $\SI{500}{\GeV}$ and $\SI{100}{\TeV}$. Since the current pattern spectra algorithm is limited in processing images with rectangular pixels, only data from the SSTs are considered in this analysis. The SST images containing the charge information, i.e. the integrated photodetector pulse, will be referred to as \textit{CTA images} in the following. \\
Some events can initiate several CTA images if the event is captured by several telescopes. The number of triggered telescopes per events, however, is not constant but varies depending on the properties of the initial particle, e.g. its type, energy and point of impact. Moreover, the construction of a CNN that can process a varying number of input images is rather challenging. As a first step towards the implementation of pattern spectra to the analysis of CTA data, we constructed a single CTA image for each event by adding up the individual pixel values of each CTA image of each event. This certainly reduces the performance of the analysis but we adopt this strategy to simplify our proof of concept work.  

\section{Analysis}
\begin{figure}[h]
    \centering
    \includegraphics[trim={0 2cm 0 2cm},clip]{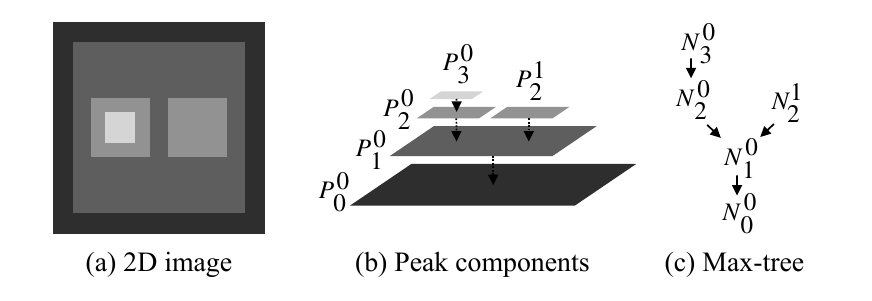}
    \caption{Basic principle of the pattern spectra algorithm (adapted from~\cite{Urbach2007,Teeninga2016})}
    \label{fig:max_tree}
\end{figure}

\noindent
The pattern spectra algorithm detects features within the CTA image and classifies these features based on their size and shape attribute. The basic working principle of the algorithm is shown in Figure~\ref{fig:max_tree}. Figure~\ref{fig:max_tree} (a) shows an example of a 2D greyscale image $f$, (b) the corresponding \textit{peak components} $P^k_h(f)$, which are defined as the $k$th grain of the threshold set $T_h(f)$ with grey level $h$, and (c) its \textit{Max-tree} composed of \textit{nodes} $N^k_h(f)$ describing the subset of the peak components $P^k_h(f)$. The size and shape attribute for each node $N^k_h(f)$ of the Max-tree is determined by computing $A(P^k_h(f))$ and $I(P^k_h(f))/A(P^k_h(f))^2$ respectively. $A(P^k_h(f))$ corresponds to the \textit{area} of the peak component and $I(P^k_h(f))$ to the \textit{moment of inertia} describing the sum of squared differences to the centre of gravity of the feature. Large (small) $A(P^k_h(f))$ values correspond to features with a large (small) size and large (small) values of $I(P^k_h(f))/A(P^k_h(f))^2$ correspond to features with a elliptical-like (circular-like) shape. For further details about the pattern spectra algorithm, we refer to Urbach et al. (2007)~\cite{Urbach2007}. 

\begin{figure}[ht]
    \centering
    \includegraphics[width=\textwidth,trim={0 0.7cm 0 0},clip]{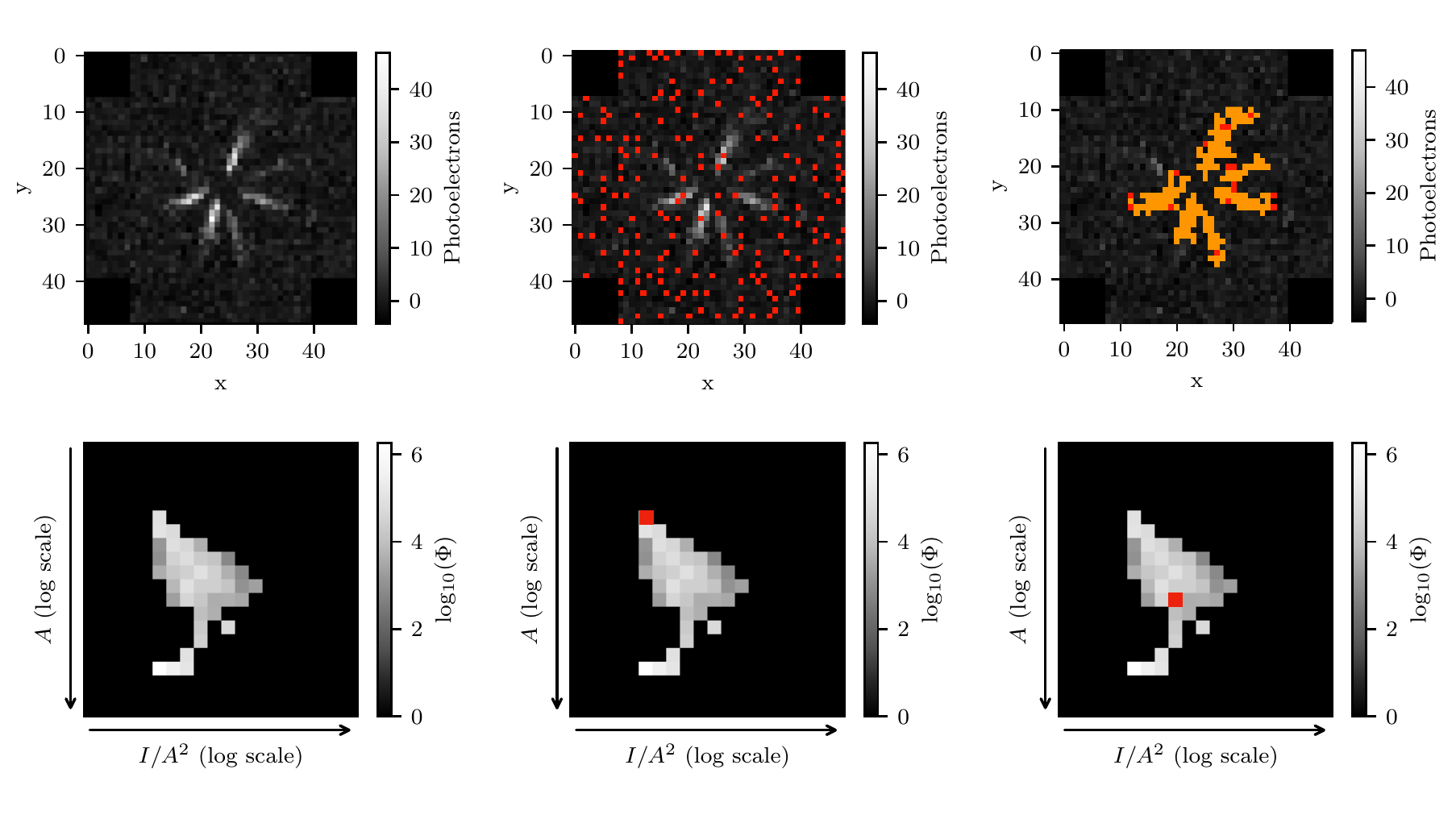}
    \caption{Top-row: CTA image example of a $\SI{1.9}{\TeV}$ gamma-ray event captured by eight SSTs with detected features highlighted in red/orange. Bottom-row: pattern spectrum extracted from the CTA image with pixel corresponding to the detected features marked in red (adapted from~\cite{Aschersleben2021}).}
    \label{fig:ps}
\end{figure}

\noindent
Figure~\ref{fig:ps} shows an example of a $\SI{1.9}{\TeV}$ gamma-ray event captured by eight SSTs (top row) and its corresponding pattern spectrum (bottom row). Each pattern spectrum pixel corresponds to a set of detected features with corresponding shape and size. Two examples of a set of detected features are shown in the second and third column of Figure~\ref{fig:ps}. The detected features in the CTA image are displayed in red/orange and the corresponding pattern spectrum pixel is marked in red. Whereas the detected features in the second column mostly consist of noise, the features in the third column consist of the Cherenkov emission initiated by the particle shower. The latter contains information about the energy and type of the initial particle which is of particular interest for this analysis.

\begin{figure}[ht]
    \centering
    \includegraphics[width=\textwidth]{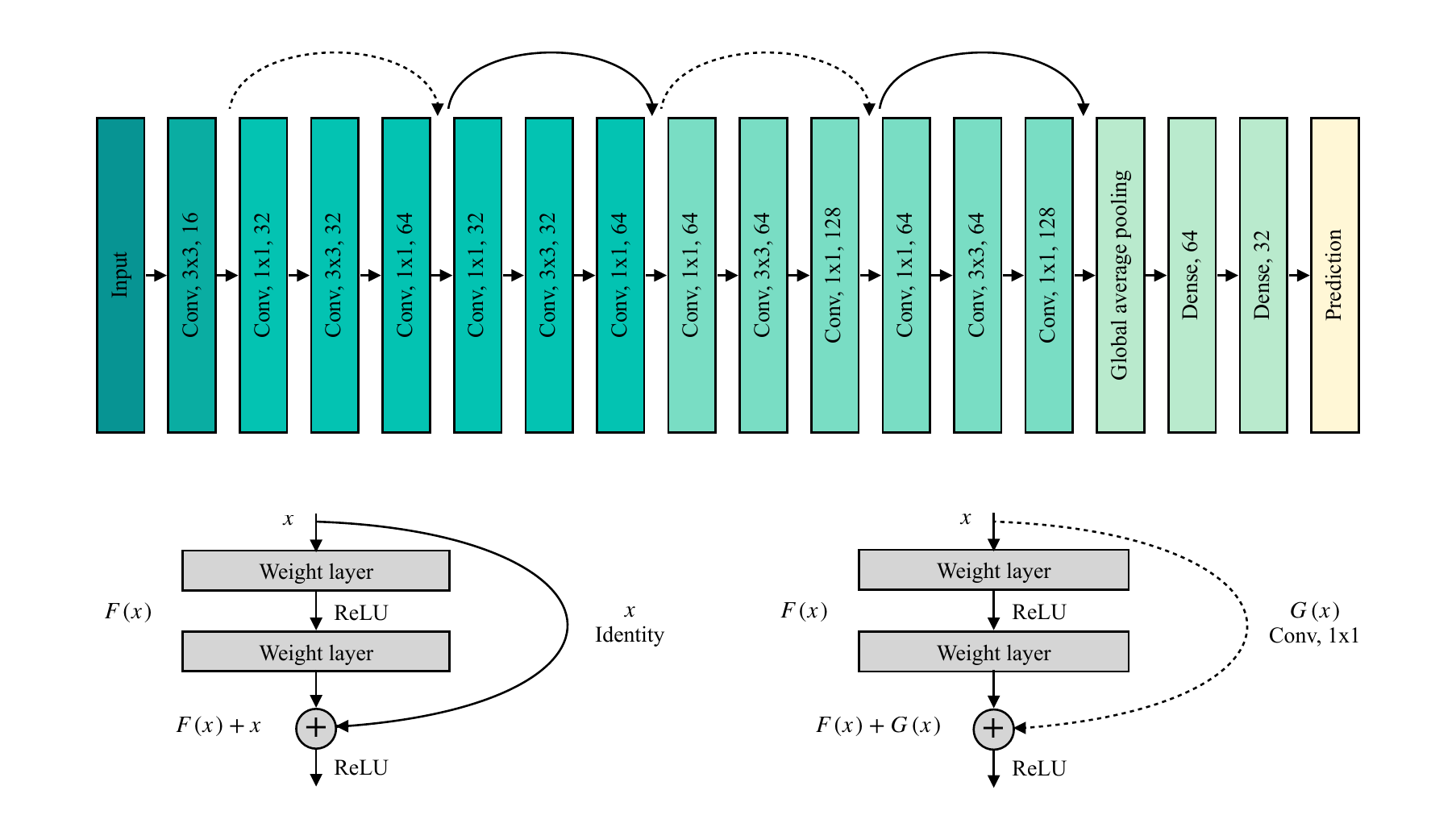}
    \caption{Top: thin residual neural network (TRN) architecture used in this work. The kernel size and number of filters for each convolutional layer and the number of neurons in each dense layer is specified in each box. The dashed and solid arrows correspond to linear (bottom left) and non-linear shortcut connections (bottom right) (adapted from~\cite{He2016}).}
    \label{fig:trn}
\end{figure}

\noindent
The architecture of our CNN is motivated by the work of Miener et al. (2021)~\cite{Miener2021b}, W. Xie et al. (2019)~\cite{Xie2019} and K. He et al. (2016)~\cite{He2016}. Two almost identical but independent CNNs are constructed for the signal-background separation and energy reconstruction task taking either the CTA images or the pattern spectra as input. We refer to our CNN as \textit{thin residual neural network} (TRN) in the following based on its rather shallow architecture compared to the residual neural network presented in K. He et al. (2016)~\cite{He2016}. The detailed architecture of the TRN is shown in Figure~\ref{fig:trn}. It consists of 13 convolutional layers, a global average pooling layer and two fully connected (dense) layers. The \textit{ReLU} activation function~\cite{Nair2010} has been applied for all convolutional and fully connected layers. One neuron is used as the output layer for the energy reconstruction task and two neurons with \textit{softmax}~\cite{Bridle1989} activation function for the signal-background separation. The TRN is constructed using Tensorflow 2.3.1~\cite{Tensorflow_bib} and Keras 2.4.3~\cite{Keras_bib} and has a total number of about 150000 trainable parameters. The data is split into $\SI{90}{\percent}$ training data, of which $\SI{10}{\percent}$ is used for validation, and $\SI{10}{\percent}$ test data. For the TRN training process, the \textit{adaptive moment} (ADAM) optimizer~\cite{Kingma2017} with a constant learning rate of $\SI{1e-3}{}$, a batch size of 32 and 30 epochs are used. The \textit{categorical cross entropy} has been applied as the loss function for the signal-background separation and the \textit{mean squared error}~\cite{Janocha2017} for the energy reconstruction. After the completion of the training, the performance of the TRN is evaluated on the test data.

\section{Results}
\subsection{Signal-background separation}
\noindent
The TRN predicts the \textit{gammaness} of the initial particle, which corresponds to a pseudo-probability of the particle being a gamma ray (photon). For a given \textit{gammaness} threshold $\alpha_g$, the \textit{true positive rate} (correctly classified photons, i.e. photons with $gammaness > \alpha_g$) and \textit{false positive rate} (misclassified protons, i.e. protons with $gammaness > \alpha_g$) are determined. The \textit{receiver operating characteristic} (ROC) curves~\cite{BRADLEY1997}, describing the \textit{true positive rate} versus the \textit{false positive rate} are shown in Figure~\ref{fig:results} (left). The area under the ROC curve (AUC) is a measure for the separation capabilities of an algorithm and calculated for both the CTA images and pattern spectra-based analyses. The TRN trained on CTA images achieves an AUC value of 0.961 which is about a factor of 1.15 higher compared to the AUC value of 0.836 achieved with the pattern spectra.

\subsection{Energy reconstruction}
\begin{figure}
    \centering
    \begin{subfigure}[b]{0.49\textwidth}
        \centering
        \includegraphics[width=\textwidth]{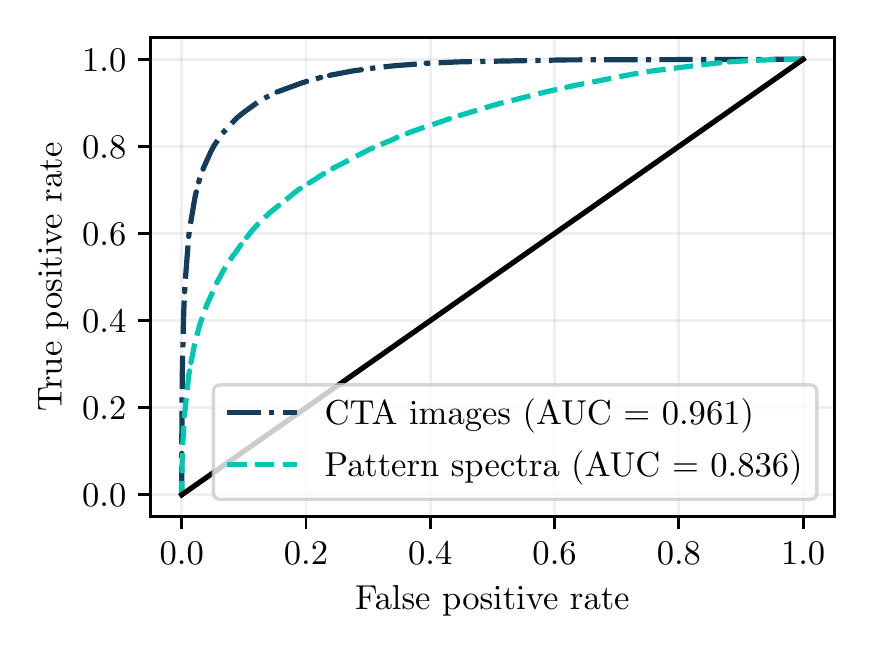}
    \end{subfigure}
    \hfill
    \begin{subfigure}[b]{0.49\textwidth}
        \centering
        \includegraphics[width=\textwidth]{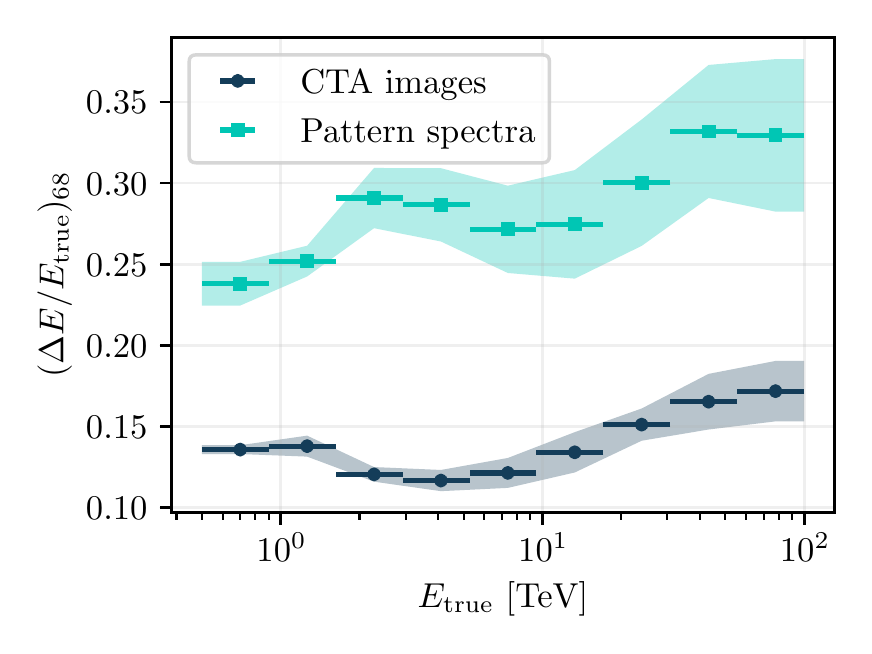}
    \end{subfigure}
       \caption{ROC curves and corresponding AUC values (left) and energy resolutions (right) obtained from the CTA images and pattern spectra-based analyses.}
       \label{fig:results}
\end{figure}

\noindent
The performance of the TRN for the energy reconstruction is evaluated by calculating the \textit{relative energy error} $\Delta E / E_{\text{true}} = (E_{\text{rec}} - E_{\text{true}}) / E_{\text{true}}$ from the reconstructed energy $E_{\text{rec}}$ obtained from the TRN and the true energy $E_{\text{true}}$. The whole $0.5-\SI{100}{\TeV}$ energy range is split logarithmically into nine bins and each event is allocated to its corresponding bin based on its true energy $E_{\text{true}}$. The distribution of the relative energy error $\Delta E / E_{\text{true}}$ is determined for each energy bin and its energy biases, i.e. the median of the relative energy error distribution, is calculated. Then, the distributions are bias-corrected by subtracting the energy bias via $(\Delta E / E_{\text{true}})_{\text{corr}} = \Delta E / E_{\text{true}} - \text{median}(\Delta E / E_{\text{true}})$. We define the energy resolution $(\Delta E / E_{\text{true}})_{68}$ as the 68th percentile of the distribution $|(\Delta E / E_{\text{true}})_{\text{corr}}|$. The training of the TRN for the energy reconstruction is performed ten times for both the CTA images and the pattern spectra in order to calculate a mean energy resolution and its standard deviation for each energy bin. The results are shown in Figure~\ref{fig:results} (right). The CTA images-based analysis outperforms the pattern spectra in each energy bin with a maximum factor of 2.5 between the two curves.

\section{Conclusions}
\noindent
In this work, we have extracted pattern spectra from CTA images and applied them on a TRN to perform signal-background separation and reconstruction of the energy of gamma rays. As shown in Table~\ref{tab:comp_power}, the training of the TRN with the pattern spectra requires about a factor of 3 less memory and is about a factor of 2.5 faster than the TRN training with CTA images. However, the CTA images-based analysis outperforms the pattern spectra in both signal-background separation and energy reconstruction capabilities. The AUC values obtained from the CTA images is about a factor of 1.15 higher than the pattern spectra value and the energy resolution obtained from the CTA images and pattern spectra differ by a maximum factor of 2.5. We therefore conclude that the pattern spectra algorithm is not able to detect all relevant features within the CTA images and/or that the size and shape attributes of the pattern spectra are not sufficient to fully describe the detected features. The pattern spectra algorithm applied on a different dataset consisting of larger images may result in a better performance. The work presented here is preliminary and results in a lower performance compared to other existing works applying CTA images on a CNN, e.g. in~\cite{Miener2021b}. The final results, including the performance on the Alpha Configuration of CTA, are discussed in~\cite{Aschersleben2022a}.

\begin{table}[ht]
    \centering
    \caption{Time and RAM required to train the TRN for signal-background separation and energy reconstruction. Since the TRN training was performed ten times for the energy reconstruction task, the mean time  $\mu_{\text{time}}$ and mean RAM $\mu_{\text{RAM}}$ have been calculated. The training was performed on a \textit{Nvidia V100 GPU}.}
    \begin{tabular}{ c | c c | c c}
        & \multicolumn{2}{c|}{$\gamma / p$-separation} & \multicolumn{2}{c}{energy reconstruction} \\ 
        &  Time &  RAM & $\mu_{\text{time}}$ & $\mu_{\text{RAM}}$ \\ \hline
        CTA images & $\SI{12.9}{\kilo \s}$ & $\SI{41.0}{\GB}$ & $\SI{29.1}{\kilo \s}$ & $\SI{100.0}{\GB}$\\ 
        pattern spectra & $\SI{5.2}{\kilo \s}$ & $\SI{13.3}{\GB}$ & $\SI{11.0}{\kilo \s}$ & $\SI{31.9}{\GB}$ \\ 
        ratio & 2.5 & 3.1 & 2.6 & 3.1 
    \end{tabular}
    \label{tab:comp_power} 
\end{table}

\section*{Acknowledgments}
\noindent
This work was conducted in the context of the CTA Consortium and CTA Observatory. We gratefully acknowledge financial support from the agencies and organizations listed at \href{http://www.cta-observatory.org/consortium_acknowledgments}{http://www.cta-observatory.org/consortium$\underline{\enskip}$acknowledgments}. We would like to thank the Centre for Information Technology of the University of Groningen for their support and for providing access to the Peregrine high performance computing cluster.


\small
\bibliography{skeleton}{}
\bibliographystyle{IEEEtran}

\section*{Full Authors List: The Cherenkov Telescope Array Consortium}


\scriptsize
\noindent
H.~Abdalla\textsuperscript{1}, H.~Abe\textsuperscript{2},
S.~Abe\textsuperscript{2}, A.~Abusleme\textsuperscript{3},
F.~Acero\textsuperscript{4}, A.~Acharyya\textsuperscript{5}, V.~Acín
Portella\textsuperscript{6}, K.~Ackley\textsuperscript{7},
R.~Adam\textsuperscript{8}, C.~Adams\textsuperscript{9}
% ...
\\

\noindent
1 : Centre for Space Research, North-West University, Potchefstroom, 2520, South Africa

\noindent
2 : Institute for Cosmic Ray Research, University of Tokyo, 5-1-5, Kashiwa-no-ha, Kashiwa, Chiba 277-8582, Japan

\noindent
3 : Pontificia Universidad Católica de Chile, Av. Libertador Bernardo O'Higgins 340, Santiago, Chile

\noindent
4 : AIM, CEA, CNRS, Université Paris-Saclay, Université Paris Diderot, Sorbonne Paris Cité, CEA Paris-Saclay, IRFU/DAp, Bat 709, Orme des Merisiers, 91191 Gif-sur-Yvette, France

\noindent
5 : Centre for Advanced Instrumentation, Dept. of Physics, Durham University, South Road, Durham DH1 3LE, United Kingdom

\noindent
6 : Port d'Informació Científica, Edifici D, Carrer de l'Albareda, 08193 Bellaterrra (Cerdanyola del Vallès), Spain

\noindent
7 : School of Physics and Astronomy, Monash University, Melbourne, Victoria 3800, Australia

\noindent
8 : Laboratoire Leprince-Ringuet, École Polytechnique (UMR 7638, CNRS/IN2P3, Institut Polytechnique de Paris), 91128 Palaiseau, France

\noindent
9 : Department of Physics, Columbia University, 538 West 120th Street, New York, NY 10027, USA

\noindent
% ...



\end{document}